\documentclass[prl,preprint,12pt,showpacs,showkeys,preprintnumbers,amsmath,amssymb,unsortedaddress,aps,longbibliography]{revtex4-1}

\usepackage{graphicx}
\usepackage{dcolumn}
\usepackage{bm}
\usepackage{amsthm}
\usepackage{amsmath}
\usepackage{amssymb}

\begin{document}

\title{Microstructural-defect-induced Dzyaloshinskii-Moriya interaction}

\author{Andreas Michels}\email{Corresponding author. andreas.michels@uni.lu}
\author{Denis Mettus}
\author{Ivan Titov}
\author{Artem Malyeyev}
\author{Mathias Bersweiler}
\author{Philipp Bender}
\author{Inma Peral}
\affiliation{Physics and Materials Science Research Unit, University of Luxembourg, 162A~Avenue de la Fa\"iencerie, L-1511 Luxembourg, Grand Duchy of Luxembourg}
\author{Rainer Birringer}
\affiliation{Experimentalphysik, Universit\"at des Saarlandes, D-66041 Saarbr\"ucken, Germany}
\author{Yifan Quan}
\author{Patrick Hautle}
\author{Joachim Kohlbrecher}
\affiliation{Paul Scherrer Institute, CH-5232~Villigen PSI, Switzerland}
\author{Dirk Honecker}
\affiliation{Institut Laue-Langevin, 71~avenue des Martyrs, CS20156, F-38042 Grenoble Cedex~9, France}
\author{Jes\'us Rodr\'iguez Fern\'andez}
\author{Luis Fern\'andez Barqu\'in}
\address{CITIMAC, Universidad de Cantabria, 39005 Santander, Spain}
\author{Konstantin L.\ Metlov}
\affiliation{Donetsk Institute for Physics and Technology, Rosa Luxembourg Str.~72, Donetsk, 83114, Ukraine}


\begin{abstract}
\bf The antisymmetric Dzyaloshinskii-Moriya interaction (DMI) plays a decisive role for the stabilization and control of chirality of skyrmion textures in various magnetic systems exhibiting a noncentrosymmetric crystal structure. A less studied aspect of the DMI is that this interaction is believed to be operative in the vicinity of lattice imperfections in crystalline magnetic materials, due to the local structural inversion symmetry breaking. If this scenario leads to an effect of sizable magnitude, it implies that the DMI introduces chirality into a very large class of magnetic materials---defect-rich systems such as polycrystalline magnets. Here, we show experimentally that the microstructural-defect-induced DMI gives rise to a polarization-dependent asymmetric term in the small-angle neutron scattering (SANS) cross section of polycrystalline ferromagnets with a centrosymmetric crystal structure. The results are supported by theoretical predictions using the continuum theory of micromagnetics. This effect, conjectured already by Arrott in 1963, is demonstrated for \textit{nanocrystalline} terbium and holmium (with a large grain-boundary density), and for \textit{mechanically-deformed} microcrystalline cobalt (with a large dislocation density). Analysis of the scattering asymmetry allows one to determine the defect-induced DMI constant, $\mathbf{\textit{D} = 0.45 \pm 0.07 \, \mathrm{\mathbf{mJ/m^2}}}$ for Tb at $\mathbf{100 \, \mathrm{\mathbf{K}}}$. Our study proves the \textit{generic} relevance of the DMI for the magnetic microstructure of defect-rich ferromagnets with vanishing intrinsic DMI. Polarized SANS is decisive for disclosing the signature of the defect-induced DMI, which is related to the unique dependence of the polarized SANS cross section on the chiral interactions. The findings open up the way to study defect-induced skyrmionic magnetization textures in disordered materials.
\end{abstract}

\maketitle

The Dzyaloshinskii-Moriya interaction (DMI) \cite{dzya58,moriya60} has recently moved again into the focus of condensed-matter research in the context of numerous observations of skyrmion lattices in various magnetic materials (see, e.g., Refs.~\onlinecite{pflei2009,pflei2010,Muenzer:10,Yu:10,Yu:11,Heinze2011,Adams:12,Seki:12,milde2013,Nagaosa2013,Tokunaga:15,Boulle2016,nayak2017,kakurai2017,chacon2018} and references therein). The origin of the DMI is related to relativistic spin-orbit coupling, and in inversion asymmetric crystal-field environments it gives rise to antisymmetric magnetic interactions \cite{bogdanov89,bogdanov94,bogdanov2005,rossler2006}. In most of the studies published so far, the DMI is the essential ingredient for the stabilization of various types of skyrmion textures, and its origin is related to the noncentrosymmetric crystal structures of the materials under study, or to the breaking of structural inversion symmetry at the interfaces in ultrathin film architectures (e.g., Refs.~\cite{Heinze2011,Yu:11,Boulle2016}). 

However, it has already been conjectured by Arrott in 1963 that lattice imperfections in the microstructure of ferromagnetic and antiferromagnetic materials are accompanied by the presence of \textit{local} chiral DMI couplings due to the breaking of inversion symmetry at defect sites~\cite{arrott1963}. Arrott suggested that the DMI is present in the vicinity of any lattice defect and that it gives rise to inhomogeneous magnetization states: for two magnetic ions which are ferromagnetically coupled by the isotropic exchange interaction, the DMI, when acting on the exchange path, produces an antiferromagnetic component, while for the two ions being antiferromagnetically aligned by isotropic exchange, the DMI causes a ferromagnetic component. In a sense, microstructural defects are supposed to act as a source of additional local chiral interactions, similar to the above mentioned \textit{intrinsic} DMI in noncentrosymmetric crystal structures. Therefore, if the above sketched scenario is valid, it is important to realize that the DMI is generally present in defect-rich magnetic materials, even in highly symmetric centrosymmetric lattices, where the ``usual'' intrinsic DMI term vanishes.

Research along these lines has previously been conducted by \textcite{Fedorov1997}, who studied the impact of torsional-strain-induced DMI couplings near dislocations on the helix domain populations in Ho metal. Similarly, \textcite{lott08} investigated the field-induced chirality in the helix structure of Dy/Y multilayer films and provided evidence for interface-induced DMI. \textcite{faehnle2010} combined ab-initio density functional electron theory with a micromagnetic model to study the DMI vectors arising from a fabrication-induced perpendicular strain gradient in a film of bcc Fe. \textcite{butenko2013} have developed a micromagnetic model for dislocation-induced DMI couplings. These authors considered a disk-like film element with a screw dislocation at its center and showed that the associated defect-induced DMI leads to a chirality selection of the vortex state.

Supported by theory \cite{michelsPRB2016}, we provide here experimental evidence for the \textit{generic} impact of the DMI on the spin microstructure of polycrystalline defect-rich ferromagnets. Examples for such systems are nanomagnets, which are characterized by a large volume fraction of internal interfaces (e.g., grain boundaries), and mechanically-deformed metals containing a large density of dislocations. In the vicinity of both types of lattice imperfections---interfaces and dislocations---inversion symmetry is likely to be broken, so that the DMI may be operative. Defect-related DMI is therefore expected to manifest in other measurements as well (e.g., magnetization data), however, the technique of polarized small-angle neutron scattering (SANS) is probably the only one which is able to directly disclose its signature: this is brought about by the unique dependence of the polarized SANS cross section on the chiral interactions. In this paper we report the results of polarized SANS experiments on nanocrystalline Tb and Ho, and on mechanically-deformed Co (see Methods section), all of which have a centrosymmetric crystal structure in the single-crystalline ground state.

Based on micromagnetic theory using the DMI energy of \textit{cubic} symmetry (e.g., \cite{rossler2006}), 
\begin{equation}
\label{edmi}
E_{\mathrm{DMI}} = \frac{D}{M_s^2} \int_V \mathbf{M} \cdot (\nabla \times \mathbf{M}) dV ,
\end{equation}
we have theoretically investigated in Ref.~\onlinecite{michelsPRB2016} the impact of the DMI on the magnetization distribution $\mathbf{M}(\mathbf{r})$ and on the ensuing magnetic SANS cross section. In addition to the DMI energy, we have taken into account the energies due to the isotropic exchange interaction, magnetic anisotropy, and the magnetostatic interaction. The central prediction is that the difference $\Delta \Sigma$ between the polarized ``spin-up'' and ``spin-down'' SANS cross sections (for the scattering geometry with the incoming neutron beam perpendicular to the applied magnetic field) is proportional to the so-called chiral function $2 \imath \chi(\mathbf{q})$, which contains the lattice-defect-induced effect related to the DMI (compare Eq.~(\ref{sanspolperpdiff}) in the Methods section):
\begin{widetext}
\begin{equation}
\label{ffinal}
2 \imath \chi(\mathbf{q}) = - \frac{2 \widetilde{H}^2_p p^3 \left( 2 + p \sin^2\theta \right) l_D q \cos^3\theta + 4 \widetilde{M}_z^2 p (1 + p)^2 l_D q \sin^2\theta \cos\theta}{\left( 1 + p \sin^2\theta - p^2 l_D^2 q^2 \cos^2\theta \right)^2} ,
\end{equation}
\end{widetext}
where $\widetilde{H}^2_p(q \xi_H)$ denotes the anisotropy-field Fourier coefficient, and $\widetilde{M}^2_z(q \xi_M)$ is the Fourier coefficient of the longitudinal magnetization. These functions characterize the strength and spatial structure of, respectively, the magnetic anisotropy field $\mathbf{H}_p(\mathbf{r})$ (with correlation length $\xi_H$) and of the \textit{local} saturation magnetization $M_s(\mathbf{r})$ (with correlation length $\xi_M$); $p = p(q, H_i) = M_s/[H_i (1 + l_H^2 q^2)]$ is a known function of $q$ and of the internal magnetic field $H_i = H_0 - N M_s$ ($N$:~demagnetizing factor); $l_H(H_i) = \sqrt{2 A / (\mu_0 M_s H_i)}$ and $l_D = 2 D / (\mu_0 M_s^2)$ represent micromagnetic length scales which characterize, respectively, the size of inhomogeneously magnetized regions around defects and the range of the DMI ($A$:~exchange-stiffness constant; $D$:~DMI constant; $\theta = \angle (\mathbf{q}, \mathbf{H}_0)$, compare Fig.~\ref{fig1}). The negative of the function $2 \imath \chi(\mathbf{q})$ is plotted in Fig.~6 in \cite{michelsPRB2016}; there, it is seen that $2 \imath \chi(\mathbf{q})$ is asymmetric in $\mathbf{q}$, which is due to the defect-induced DMI: at small fields, when the anisotropy term $\propto \widetilde{H}^2_p$ in the numerator of Eq.~(\ref{ffinal}) dominates, two extrema parallel and antiparallel to the field axis are observed, whereas at larger fields, when the magnetostatic term $\propto \widetilde{M}_z^2$ comes into play, additional maxima and minima appear approximately along the detector diagonals. Moreover, we note that $\chi(\mathbf{q})$ vanishes for purely real magnetization Fourier components (irrespective of the value of the applied magnetic field), and at complete magnetic saturation when $\mathbf{M}(\mathbf{r}) = \{0, 0, M_z = M_s\}$ (compare Eq.~(\ref{chiral}) in Methods section).

Evaluating Eq.~(\ref{ffinal}) along the horizontal direction ($\theta = 0^{\circ}$ and $\theta = 180^{\circ}$) and assuming that $\widetilde{H}^2_p$ depends only on the magnitude of $\mathbf{q}$, we have (independent of $\widetilde{M}_z^2$)
\begin{equation}
\label{fffinal}
2 \imath \chi(q, H_i) = \mp \frac{4 \widetilde{H}^2_p p^3 l_D q}{\left( 1 - p^2 l_D^2 q^2 \right)^2} ,
\end{equation}
which can be used to analyze experimental data. Note that $\chi(\mathbf{q}) = 0$ for $\theta = 90^{\circ}$, which allows its clear separation from the nuclear-magnetic interference term $\propto \sin^2\theta$ in Eq.~(\ref{sanspolperpdiff}) (provided that both $\widetilde{N}$ and $\widetilde{M}_z$ are isotropic). In the analysis below, we assume that the anisotropy-field Fourier coefficient is described by a squared Lorentzian, $\widetilde{H}^2_p(q \xi_H) = \langle H_p^2 \rangle / \left( 1 + \xi_H^2 q^2 \right)^2$, where $\langle H_p^2 \rangle$ is the mean-square anisotropy field, and $\xi_H$ denotes the correlation length of the anisotropy field. For an idealized nanocrystalline ferromagnet, where each grain is a single crystal and the anisotropy field jumps randomly in direction at grain boundaries due to the changing set of crystallographic easy axes, the correlation length $\xi_H$ is expected to be related to the average grain size. Mechanical deformation of Co results in an increased density of dislocation structures, which, by means of magnetoelastic coupling, gives rise to inhomogeneous magnetization states and to a concomitant magnetic SANS contrast~\cite{kronfahn03}; in this case, $\xi_H$ characterizes the average extension of the dislocation.

Before discussing the experimental results we find it appropriate to pause and to briefly examine how defects in the microstructure of a magnetic material (e.g., vacancies, dislocations, grain boundaries, pores) are related to the diffuse magnetic SANS cross section. The following considerations are restricted to the mesoscopic length scale which is accessible by the conventional SANS method ($\sim 1-300 \, \mathrm{nm}$), as it is suitable for a description within the micromagnetic continuum picture. The mechanisms by which nanoscale spin disorder in magnetic materials is generated are essentially related to (i)~spatial variations in the magnetic anisotropy field $\mathbf{H}_p(\mathbf{r})$, and to (ii)~spatial variations in the magnetic materials parameters, most notably the local saturation magnetization $M_s(\mathbf{r})$. Both sources of magnetization inhomogeneities---modeling the effect of the defects---are taken into account in the magnetic SANS theory~\cite{michelsPRB2016}. To be more specific, forces due to the distortion of the crystal lattice in the vicinity of a microstructural defect tend to rotate the local magnetization vector field $\mathbf{M}(\mathbf{r})$ along the main axes of the system of internal stresses (so-called magnetoelastic coupling), while magnetocrystalline anisotropy tries to pull the magnetic moments along the principal axes of the crystal \cite{brown40}. Likewise, nanoscale spatial variations of the saturation magnetization, exchange, or anisotropy constants (e.g., at internal interfaces in a magnetic nanocomposite or in a nanoporous ferromagnet) give rise to inhomogenous magnetization states, which represent a contrast for magnetic SANS. It is also important to emphasize that the adjustment of the magnetization along the respective local ``easy'' axes does not occur abruptly, i.e., on a scale of the interatomic spacing, but requires a more extended range. This is a consequence of the quantum-mechanical exchange interaction, which spreads out local perturbations in the magnetization (at the defect core) over larger distances. The size of such spin inhomogeneities is characterized by the micromagnetic exchange length $l_H \propto H_i^{-1/2}$~\cite{brown40,seeger61}, which varies continuously with the applied field and takes on values between about $1-100 \, \mathrm{nm}$, a size regime which is routinely accessible by SANS~(e.g., \cite{michels03prl,mirebeau2018}). As we will show below, the defect-related DMI introduces chirality into the system and renders the spin distribution around the defect cores asymmetric.

\begin{figure}[tb!]
\centering
\resizebox{0.95\columnwidth}{!}{\includegraphics{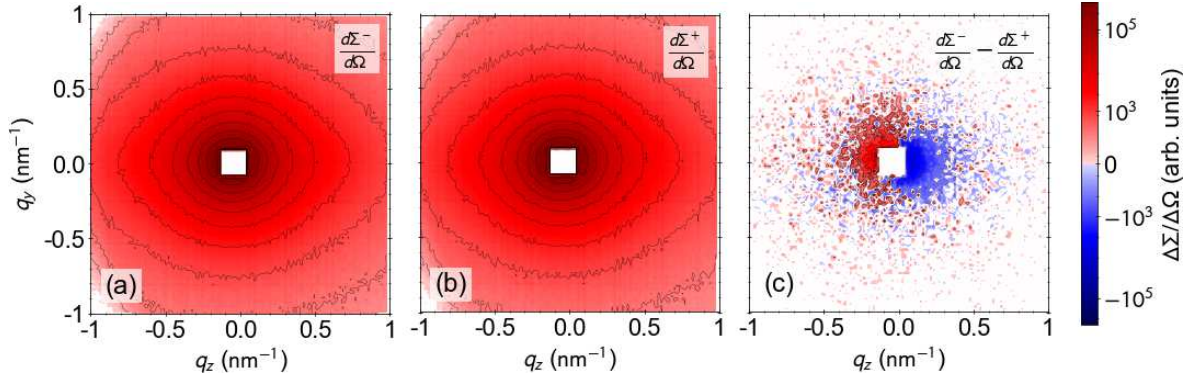}}
\caption{DMI asymmetry in nanocrystalline Tb. Shown are polarized SANS results of nanocrystalline Tb at $100 \, \mathrm{K}$ and $5 \, \mathrm{T}$ ($\mathbf{H}_0 \parallel \mathbf{e}_z$). (a)~Two-dimensional flipper-on SANS cross section $d \Sigma^- / d \Omega$; (b)~flipper-off SANS cross section $d \Sigma^+ / d \Omega$; (c)~$\Delta\Sigma = d \Sigma^- / d \Omega - d \Sigma^+ / d \Omega$. The difference signal amounts to $\sim 8.5 \, \%$ of the $d \Sigma^{\pm} / d \Omega$.}
\label{fig2}
\end{figure}

Figure~\ref{fig2} depicts the results of polarized SANS experiments on nanocrystalline Tb (average crystallite size: $L = 40 \, \mathrm{nm}$ \cite{smdmi2018}) in the ferromagnetic state at $T = 100 \, \mathrm{K}$ and at an applied magnetic field of $\mu_0 H_0 = 5 \, \mathrm{T}$ \cite{tbpropcomment}. We emphasize that it is a well-documented result in the literature that the magnetic properties of nanocrystalline heavy rare-earth metals are strongly affected by a small crystallize size, i.e., by an associated large volume fraction of interfaces \cite{dami02,weissm04a,michels11a,dobrichprb2012,szary2016} (see also Fig.~2 in \cite{smdmi2018}). Both spin-resolved data sets $d \Sigma^- / d \Omega$ [Fig.~\ref{fig2}(a)] and $d \Sigma^+ / d \Omega$ [Fig.~\ref{fig2}(b)] are characterized by a maximum of the scattering intensity along the horizontal applied-field direction. This is the signature of spin-misalignment scattering due to the presence of transversal magnetization components [cf.\ the term $|\widetilde{M}_y|^2 \cos^2\theta$ in Eq.~(\ref{sanspolperp})]. The difference between the two spin-resolved SANS cross sections, $\Delta\Sigma = d \Sigma^- / d \Omega - d \Sigma^+ / d \Omega$ [Fig.~\ref{fig2}(c)], clearly exhibits an \textit{asymmetric} contribution related to the chiral function [compare Eq.~(\ref{sanspolperpdiff})]. The asymmetry is most pronounced along the horizontal direction, which by comparison to the theoretical prediction [Eq.~(\ref{ffinal})] can be attributed to the anisotropy-field term $\propto - \widetilde{H}^2_p \cos^3\theta$. Data taken at smaller momentum transfers additionally show the ``usual'' \textit{symmetric} $\sin^2\theta$-type anisotropy (with maxima at $\theta = 90^{\circ}$ and $\theta = 270^{\circ}$), which is due to the polarization-dependent nuclear-magnetic interference term in the SANS cross section (see Fig.~4 in \cite{smdmi2018}).

\begin{figure}[tb!]
\centering
\resizebox{0.50\columnwidth}{!}{\includegraphics{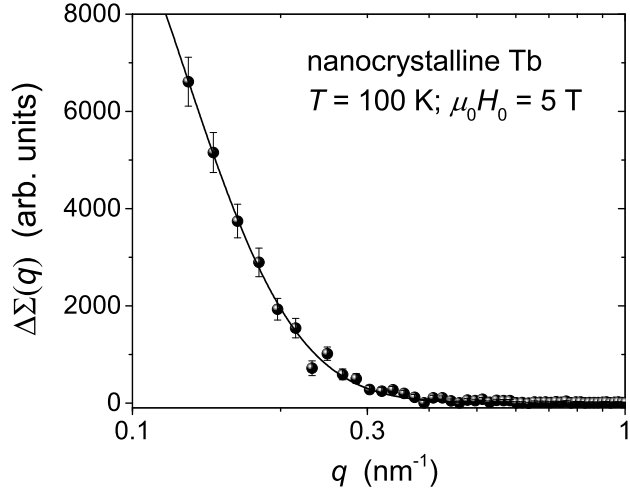}}
\caption{Determination of the defect-induced DMI strength. ($\bullet$)~Azimuthally-averaged $\Delta\Sigma(q)$ of the data shown in Fig.~\ref{fig2}(c) (log-linear scale). For the data analysis the arithmetic mean of both horizontal $\pm 10^{\circ}$ sector-averaged $\Delta\Sigma$ branches were employed (taking into account their different signs). Error bars represent $\pm 1$ standard deviation. Solid line:~fit to Eq.~(\ref{fffinal}) with ``$+$'' sign.}
\label{fig3}
\end{figure}

Figure~\ref{fig3} displays the angular average of the $\Delta\Sigma(\mathbf{q})$ data from Fig.~\ref{fig2}(c) along the horizontal direction. By performing a weighted nonlinear least-squares fit of the resulting data to Eq.~(\ref{fffinal}) (solid line in Fig.~\ref{fig3}), the following parameters are obtained: $D = 0.45 \pm 0.07 \, \mathrm{mJ/m^2}$; $A = 8.2 \pm 2.0 \times 10^{-11} \, \mathrm{J/m}$; $\xi_H = 7.6 \pm 1.0 \, \mathrm{nm}$. The value for $D$ is comparable to bulk DMI values (e.g., \cite{iguchi2015,seki2016}), the effective $A$-value is slightly increased but still within the range of experimental data \cite{kronfahn03,weissm04a}, while the correlation length $\xi_H$ of the anisotropy-field is smaller than the average crystallite size of the Tb sample ($L = 40 \, \mathrm{nm}$ \cite{smdmi2018}). The latter finding indicates that there is a significant spin disorder within the grains. This is in agreement with the results of an earlier unpolarized SANS study~\cite{weissm04a}, which has found that up to applied fields of several Tesla the magnetization remains ``locked in'' to the basal planes of the hcp crystal lattice of each individual crystallite, but that the in-plane orientation of the spins is highly nonuniform within each grain.

\begin{figure}[tb!]
\centering
\resizebox{0.80\columnwidth}{!}{\includegraphics{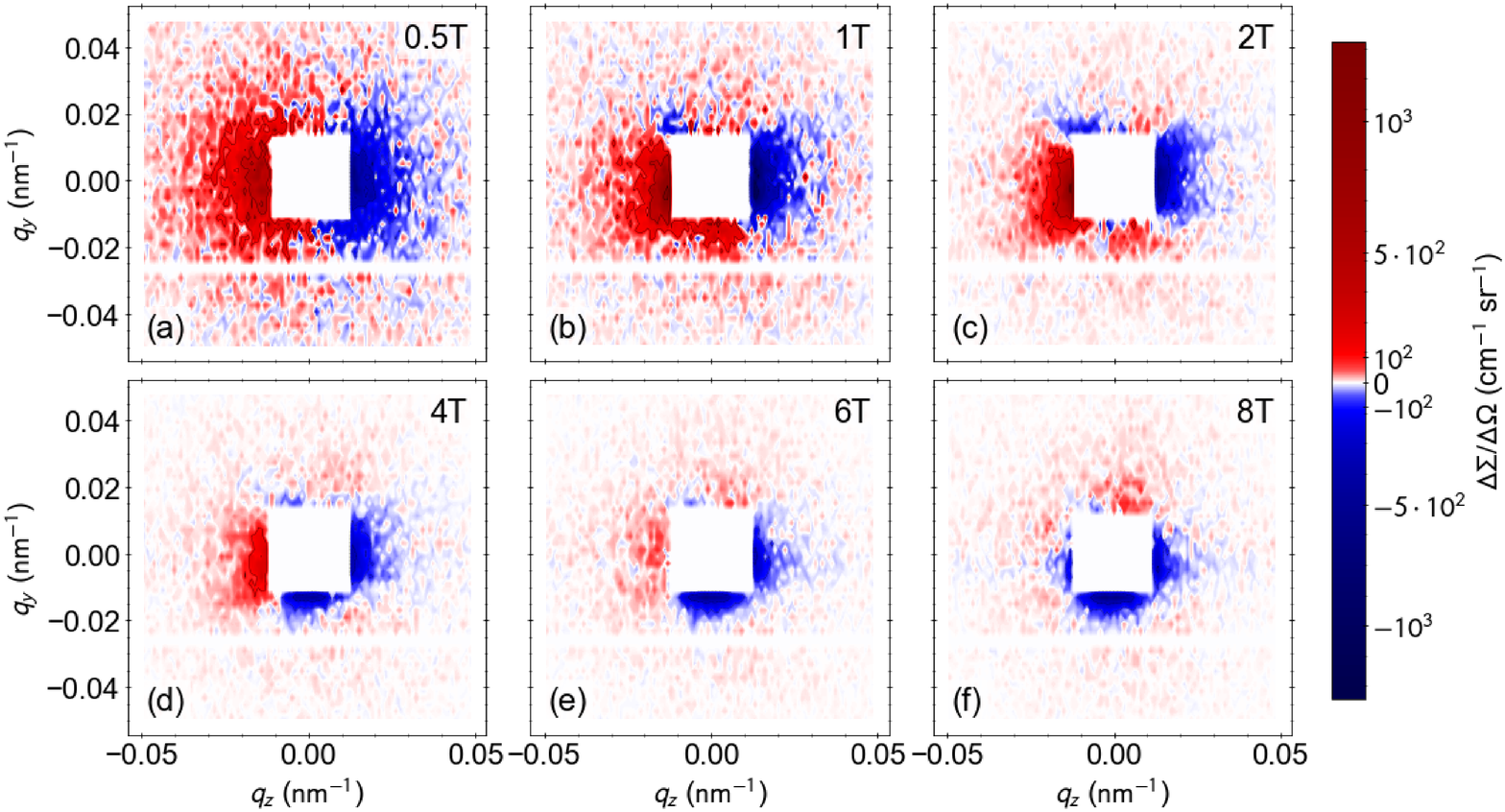}}
\resizebox{0.80\columnwidth}{!}{\includegraphics{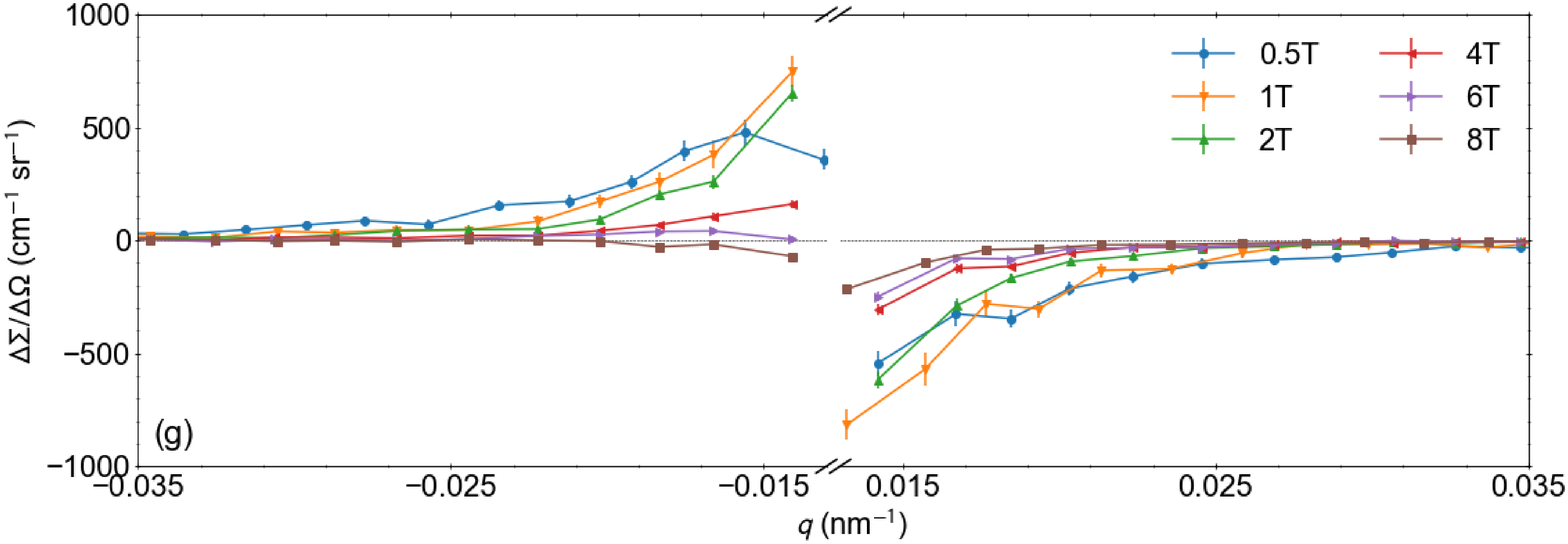}}
\caption{Field dependence of DMI asymmetry in mechanically-deformed Co. Displayed are polarized SANS results of mechanically-deformed Co at $T = 300 \, \mathrm{K}$ and at a series of applied magnetic fields (see insets). $\mathbf{H}_0 \parallel \mathbf{e}_z$ and parallel to the rolling direction. (a)-(f)~Two-dimensional flipper-on minus flipper-off data $\Delta\Sigma(\mathbf{q})$. (g)~Azimuthally-averaged $\Delta\Sigma(q)$ ($\pm 10^{\circ}$ horizontal sector averages). Error bars represent $\pm 1$ standard deviation.}
\label{fig4}
\end{figure}

Magnetic-field-dependent $\Delta\Sigma$ data of the deformed Co sample at room temperature are shown in Fig.~\ref{fig4} with the magnetic field applied parallel to the rolling direction. Inspection of the Co magnetization curves (Fig.~3 in \cite{smdmi2018}) clearly shows that all of the displayed field values fall into the approach-to-saturation regime, where the predictions of the micromagnetic SANS theory are valid~\cite{michelsPRB2016}. The results confirm, in agreement with the theory, that with increasing field the asymmetry decreases and eventually vanishes for fields in excess of $\sim 8 \, \mathrm{T}$; note that $\chi \rightarrow 0$ for $H_i \rightarrow \infty$ (compare Eq.~(\ref{chiral}) in the Methods section).

As shown in the Supplemental Material (Fig.~5 in \cite{smdmi2018}), polarized SANS data on nanocrystalline Ho (grain size:~$33 \pm 3 \, \mathrm{nm}$) at $1.8 \, \mathrm{K}$ and $5 \, \mathrm{T}$ also reveal the scattering asymmetry. At this temperature Ho exhibits a conical ferromagnetic spin structure~\cite{legvold80}. This result, together with the data on Tb and Co, underlines the generic character of the DMI asymmetry for polycrystalline magnetic materials having a centrosymmetric crystal structure.

\begin{figure}[tb!]
\centering
\includegraphics[width=0.60\columnwidth]{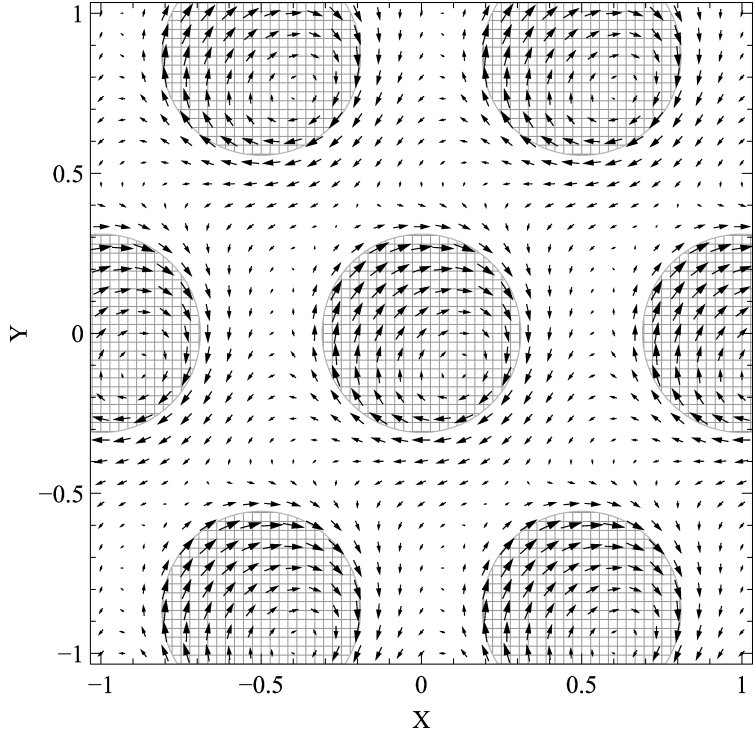}
\caption{\label{fig5} Real-space image of the spin structure around defects in the presence of DMI. Shown is the transversal magnetization distribution around a periodic columnar array of circular defects, obtained from the solution of the micromagnetic problem in Ref.~\onlinecite{michelsPRB2016}. The defects (shaded areas) are regions in space where the local saturation magnetization is increased and where the magnetic anisotropy field is nonzero. They form a hexagonal lattice in the $x$-$y$ plane and are uniform in the $z$-direction ($\mathbf{H}_0 \parallel \mathbf{e}_z$). The exchange length $l_M = 0.2$ and the DMI length $l_D = 1$ are in units of the $x$-lattice period. The internal field is $H_i = M_s$ and the anisotropy field is $h_x = h_y = 0.2$. The defect strength is absorbed as a factor into the arrow length. See the Supplemental Material for further information~\cite{smdmi2018}.}
\end{figure}

In the following we will discuss the possible mesoscale real-space spin structure which lies at the heart of the present SANS data. Recently, \textcite{mirebeau2018} have also provided a SANS study, albeit using unpolarized neutrons, of a disordered magnetic material, the reentrant spin glass $\mathrm{Ni}_{0.81}\mathrm{Mn}_{0.19}$. Supported by the results of Monte Carlo simulations of a two-dimensional lattice structure with competing ferromagnetic and antiferromagnetic exchange interactions, these authors have suggested the existence of vortex-like chiral spin structures. While vortices are ubiquitous in magnetism and can very well be at the origin of the underlying mesoscopic spin structure, our data directly show that there is a general symmetry breaking in the samples (besides the one induced by the external field) resulting in a dominant overall chirality of the spin texture. In the case of the Co sample the reason for such a symmetry breaking is obviously related to the mechanical deformation (similar to the work by \textcite{Fedorov1997}). Regarding the nondeformed samples (Tb and Ho), the vorticity in the magnetization texture appears only because the samples are intrinsically inhomogeneous. The shape of the chiral structures in our samples is not an equilibrium property of the uniform magnet, but directly reflects the shape of the magnetic inhomogeneities. This is demonstrated in Fig.~\ref{fig5}, which displays the numerically computed real-space spin structure around defects in the presence of DMI (see \cite{smdmi2018} for details). An asymmetry in the transversal spin configuration clearly becomes evident, which is a consequence of the DMI: setting $l_D \propto D = 0$ produces a symmetric magnetization pattern with a maximum transversal spin deviation located at the centers of the defects (due to the perturbing effect of the magnetic anisotropy field). By contrast, for $l_D \neq 0$, it is seen that the chirality in the magnetization texture manifests itself at/near the boundaries of the inhomogeneities (boundaries of the shaded areas in Fig.~\ref{fig5}), precisely where one would expect the antisymmetric DMI to appear. Thus, the value of the DMI constant obtained from the SANS experiment should mainly reflect this emergent DMI strength.

To summarize, using polarized small-angle neutron scattering (SANS) and micromagnetic theory we have provided evidence that the Dzyaloshinskii-Moriya interaction (DMI) which is due to the lack of structural inversion symmetry at microstructural-defect sites gives rise to an asymmetry in the polarized SANS cross section. This has been demonstrated for the cases of nanocrystalline terbium and holmium, and for mechanically-deformed microcrystalline cobalt, all of which have a centrosymmetric crystal structure. To the best of our knowledge, our study proves for the first time the generic relevance of the DMI for the spin microstructure of defect-rich ferromagnets, where this interaction introduces chirality into the system. Consequently, the effect should also show up in other magnetic measurements as well, for instance, in the approach-to-saturation regime of a magnetization curve. However, due to the unique dependence of the polarized neutron scattering cross section on the chiral function, the SANS method is particularly powerful for revealing the fingerprint of the defect-induced DMI. An analytical theoretical framework for determining the DMI constant, which is a key parameter in the search for skyrmion-hosting materials and for the elucidation of the potential of skyrmions for high-density magnetic data storage, has been established. Our findings may provide an impetus towards the investigation of defect-induced local chiral spin textures in polycrystalline and disordered materials, such as spin glasses. Since the strength of the effect scales with the spin-orbit coupling, one may combine low-anisotropy (e.g., Gd$^{3+}$ which is a pure $S$-state ion) with high-anisotropy materials.

\section*{Methods}

\subsection*{Sample}

A circular disk-shaped nanocrystalline Tb sample with a diameter of $8 \, \mathrm{mm}$ and a thickness of $0.355 \, \mathrm{mm}$ was synthesized by the inert-gas condensation technique, as described in detail in Refs.~\onlinecite{dami02,weissm04a,michels11a,dobrichprb2012,szary2016}. The average crystallite size of the as-prepared nanocrystalline Tb sample was determined by analysis of wide-angle x-ray diffraction data and found to be $L = 40 \pm 20 \, \mathrm{nm}$~\cite{smdmi2018}. The results of the structural and magnetic characterization of the nanocrystalline inert-gas-condensed Ho sample ($L = 33 \pm 3 \, \mathrm{nm}$) are reported in Ref.~\onlinecite{szary2016}. Commercially available high-purity polycrystalline Co pieces (purity:~$99.99$$+ \, \%$) were arc-melted in an argon atmosphere. A thin disk was cut from the as-prepared sample; cold-rolling resulted in a reduction of the thickness of the disk from $0.71 \, \mathrm{mm}$ to $0.57 \, \mathrm{mm}$. In the SANS experiment, the rolling direction was oriented parallel to the applied magnetic field. Vibrating sample and SQUID magnetometry was used to measure hysteresis loops and ac~susceptibility~\cite{smdmi2018}. 

\subsection*{SANS Experiment}

\begin{figure}[tb!]
\centering
\resizebox{0.65\columnwidth}{!}{\includegraphics{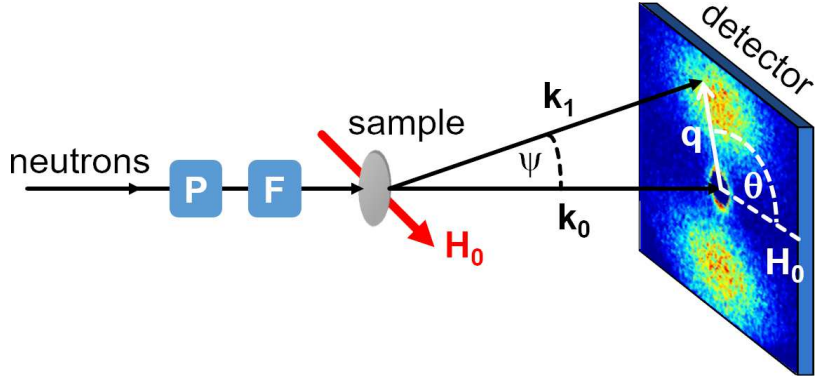}}
\caption{Schematic drawing of the scattering geometry. The scattering vector $\mathbf{q}$ is defined as $\mathbf{q} = \mathbf{k}_1 - \mathbf{k}_0$, where $\mathbf{k}_0$ and $\mathbf{k}_1$ are the wave vectors of the incident and scattered neutrons; $q = |\mathbf{q}| = (4\pi / \lambda) \sin(\psi/2)$ depends on the mean wavelength $\lambda$ of the neutrons and on the scattering angle $\psi$. Symbols ``P'' and ``F'' denote, respectively, the polarizer and the spin flipper. $\mathbf{H}_0 = H_0 \mathbf{e}_z$, $\mathbf{k}_0 = 2\pi \lambda^{-1} \mathbf{e}_x$, so that $\mathbf{q} \cong q \{ 0, \sin\theta, \cos\theta \}$ in small-angle approximation.}
\label{fig1}
\end{figure}

The neutron experiment was conducted at the D33 instrument at the Institut Laue-Langevin, Grenoble, France~\cite{dewhurst2016,illreportdmi2018}. We used polarized incident neutrons with a mean wavelength of $\lambda = 6 \, \mathrm{\AA}$ and a wavelength broadening of $\Delta \lambda / \lambda = 10 \, \%$ (FWHM); for the runs on deformed Co, the wavelength was increased to $\lambda = 12 \, \mathrm{\AA}$ ($\Delta \lambda / \lambda = 10 \, \%$) in order to access smaller momentum transfers. The external magnetic field $\mathbf{H}_0$ was provided by a cryomagnet and applied perpendicular to the wave vector $\mathbf{k}_0$ of the incident neutron beam; see Fig.~\ref{fig1} for a schematic drawing of the experimental neutron setup. The beam was polarized by a remanent FeSi supermirror transmission polarizer ($m = 3.6$), and a radio-frequency (rf) spin flipper allowed us to reverse the initial neutron polarization. The flipping efficiency of the rf flipper was $\epsilon = 99.8 \, \%$, and the polarizer efficiency was $P = 97.6 \, \%$ at $\lambda = 6 \, \mathrm{\AA}$ and $P = 95.4 \, \%$ at $\lambda = 12 \, \mathrm{\AA}$. Further neutron experiments under similar conditions have been performed at the SANS~I instrument at the Paul Scherrer Institute, Villigen, Switzerland~\cite{kohlbrecher2000}. For SANS data reduction the GRASP software package was used \cite{graspurl}.

\subsection*{Spin-polarized SANS cross section}

Neglecting nuclear spin-dependent scattering, the two half-polarized elastic differential SANS cross sections for the perpendicular scattering geometry ($\mathbf{k}_0 \perp \mathbf{H}_0$) can be expressed as~\cite{maleyev63,blume63,michels08rop,rmp2018}:
\begin{eqnarray}
\label{sanspolperp}
\frac{d \Sigma^{\pm}}{d \Omega} = \frac{8 \pi^3}{V} b_H^2 \left( b_H^{-2} |\widetilde{N}|^2 + |\widetilde{M}_x|^2 + |\widetilde{M}_y|^2 \cos^2\theta + |\widetilde{M}_z|^2 \sin^2\theta \right. \nonumber \\ \left. - (\widetilde{M}_y \widetilde{M}_z^{\ast} + \widetilde{M}_y^{\ast} \widetilde{M}_z) \sin\theta \cos\theta + (2P - 1) (2\epsilon^{\pm} - 1) b_H^{-1} (\widetilde{N} \widetilde{M}_z^{\ast} + \widetilde{N}^{\ast} \widetilde{M}_z) \sin^2\theta \right. \nonumber \\ \left. - (2P - 1) (2\epsilon^{\pm} - 1) b_H^{-1} (\widetilde{N} \widetilde{M}_y^{\ast} + \widetilde{N}^{\ast} \widetilde{M}_y) \sin\theta \cos\theta + \imath (2P - 1) (2\epsilon^{\pm} - 1) \chi \right),
\end{eqnarray}
where the chiral function $\chi(\mathbf{q})$ is given by
\begin{eqnarray}
\label{chiral}
\chi(\mathbf{q}) = \left( \widetilde{M}_x \widetilde{M}_y^{\ast} - \widetilde{M}_x^{\ast} \widetilde{M}_y \right) \cos^2\theta - \left( \widetilde{M}_x \widetilde{M}_z^{\ast} - \widetilde{M}_x^{\ast} \widetilde{M}_z \right) \sin\theta \cos\theta .
\end{eqnarray}
In Eq.~(\ref{sanspolperp}), $V$ is the scattering volume, the constant $b_H = 2.91 \times 10^8 \, \mathrm{A^{-1} m^{-1}}$ relates the atomic magnetic moment $\mu_a$ to the atomic magnetic scattering length $b_m = b_H \mu_a$, $\widetilde{N}(\mathbf{q})$ and $\widetilde{\mathbf{M}}(\mathbf{q}) = \{ \widetilde{M}_x, \widetilde{M}_y, \widetilde{M}_z \}$ denote, respectively, the Fourier transforms of the nuclear scattering-length density and of the magnetization vector field $\mathbf{M}(\mathbf{r}) = \{ M_x, M_y, M_z \}$, $\theta$ is the angle between $\mathbf{H}_0 = H_0 \mathbf{e}_z$ and $\mathbf{q}$, so that $\mathbf{q} \cong q \{ 0, \sin\theta, \cos\theta \}$ in small-angle approximation, $\imath^2 = -1$, and the asterisks ``$*$'' mark the complex-conjugated quantity. Note that $\chi = 0$ for $H_0 \rightarrow \infty$ (when $M_x = M_y = 0$) and for purely real magnetization Fourier components. The efficiency of the polarizer is denoted by $P$, and $\epsilon^{\pm}$ is the efficiency of the spin flipper ($\epsilon^{+} = 0$ for flipper off and $\epsilon^{-} = \epsilon \cong 1$ for flipper on). In our data analysis we have neglected the polarization-dependent terms $\propto \widetilde{N} \widetilde{M}_y \sin\theta \cos\theta$ in Eq.~(\ref{sanspolperp}). These contributions average out for statistically-isotropic polycrystalline magnetic materials, since there are no correlations between spatial variations in the nuclear density and in the transversal magnetization components (note that $\langle M_y \rangle = V^{-1} \int_V M_y(\mathbf{r}) dV = 0$ for such a material). The difference $\Delta\Sigma$ between flipper-on (``$-$'') and flipper-off (``$+$'') SANS cross sections reads:
\begin{eqnarray}
\label{sanspolperpdiff}
\Delta\Sigma = \frac{d \Sigma^{-}}{d \Omega} - \frac{d \Sigma^{+}}{d \Omega} = \frac{8 \pi^3}{V} b_H^2 (2P - 1) \epsilon \left[ 2 b_H^{-1} (\widetilde{N} \widetilde{M}_z^{\ast} + \widetilde{N}^{\ast} \widetilde{M}_z) \sin^2\theta + 2 \imath \chi \right] .
\end{eqnarray}

\subsection*{Influence of inelastic SANS}

Typical SANS instrumentation does not allow for energy analysis of the scattered neutrons and the measurable quantity is the energy-integrated macroscopic differential scattering cross section $d \Sigma / d \Omega$ at scattering vector $\mathbf{q}$. Since the asymmetry which we report is due to the effect of the DMI on the \textit{static} magnetic microstructure, which is probed by \textit{elastic} scattering, it is necessary to show that inelastic magnon scattering is suppressed in the small-angle regime under the conditions of our experiment. In this context we emphasize that the here reported DMI asymmetry is very much different than the so-called ``left-right'' asymmetry \cite{oko86,grigo2015}, which has its origin in inelastic spin-wave scattering. In Ref.~\cite{michels08rop} it was shown that the requirements of conservation of momentum and energy upon absorption or emission of a magnon cannot be satisfied simultaneously for any scattering vector in the small-angle regime when the applied magnetic field exceeds the critical value:
\begin{eqnarray}
\label{spinwaveeq}
\mu_0 H^{\ast}_0 \cong \frac{\hbar^4 k_0^2}{4 m^2 g \mu_B K} - \frac{\Delta}{g \mu_B} ,
\end{eqnarray}
where $k_0 = 2\pi/\lambda$, $K$ is the spin-wave stiffness constant, and $\Delta$ denotes the gap energy in the (simplified) spin-wave dispersion relation $E_q = \Delta + g \mu_B \mu_0 H_0 + K q^2$  ($\hbar$:~Planck's constant divided by $2\pi$; $m$:~neutron mass; $g$:~g-factor; $\mu_B$:~Bohr magneton). Inserting typical values of these parameters for Tb \cite{weissm04a} and Co \cite{michels08rop}, one can see that inelastic magnon scattering is suppressed at the applied-field values of the present experiment ($\mu_0 H^{\ast}_0 \cong 0.08 \, \mathrm{T}$ for Co and $\mu_0 H^{\ast}_0 \cong 0.4 \, \mathrm{T}$ for Tb at, respectively, zero gap energy).

\section*{Acknowledgements}

A.Mi.\ and D.M.\ thank the National Research Fund of Luxembourg for financial support (Project No.~INTER/DFG/12/07). This work is based on experiments performed at the Institut Laue-Langevin, Grenoble, France, and at the Swiss spallation neutron source SINQ, Paul Scherrer Institute, Villigen, Switzerland. We thank Sebastian M\"uhlbauer for the critical reading of the manuscript.

\section*{Author Contributions}

A.Mi.\ conceived and designed the study and wrote the manuscript. A.Mi., D.M., I.T., A.Ma., M.B., P.B., Y.Q., P.H., J.K., and D.H.\ performed and analyzed the polarized neutron-scattering experiments. D.M., I.T., A.Ma., I.P., R.B., J.R.F., and L.F.B.\ carried out sample synthesis as well as the magnetization and x-ray diffraction experiments. A.Mi.\ and K.L.M.\ are responsible for the micromagnetic part of the paper; K.L.M.\ performed the micromagnetic calculation of the real-space magnetization. All authors discussed the results and commented on the manuscript.

\section*{Additional Information}

Supplementary information accompanies this paper at...

\section*{Competing Interests:}

The authors declare no competing interests.


\begin{thebibliography}{52}
\bibitem [{tbp()}]{tbpropcomment}%
  \BibitemOpen
  \href@noop {} {}\bibinfo {note} {We remind that Tb is paramagnetic above the
  N$\mathrm{\acute{e}}$el point at $T_N = 230 \, \mathrm{K}$ and exhibits an
  antiferromagnetic helical state for temperatures between $219.5 \, \mathrm{K}
  < T < 230 \, \mathrm{K}$; below $T_C = 219.5 \, \mathrm{K}$ it is a
  basal-plane ferromagnet \cite{legvold80}.}\BibitemShut {Stop}%
\end{thebibliography}

%

\end{document}